*Article*

# D2WFP: A Novel Protocol for Forensically Identifying, Extracting, and Analysing Deep and Dark Web Browsing Activities

**Mohamed Chahine Ghanem*** [1&2], **Patrick Mulvihill** [1&3], **Karim Ouazzane** [1], **Ramzi Djemai** [1], **Dipo Dunsin** [1],

[1] Cyber Security Research Centre, London Metropolitan University
[2] Department of Computer Science, University of Liverpool
[3] Cyber Threat Intelligence, Grant Thornton UK LLP
* Correspondence: Dr Mohamed Chahine Ghanem ( ghanemm@staff,londonmet.ac.uk)

**Abstract:** The use of the un-indexed web, commonly known as the deep web and dark web, to commit or facilitate criminal activity has drastically increased over the past decade. The dark web is an infamously dangerous place where all kinds of criminal activities take place [1-2], despite advances in web forensics techniques, tools, and methodologies, few studies have formally tackled the dark and deep web forensics and the technical differences in terms of investigative techniques and artefacts identification and extraction. This research proposes a novel and comprehensive protocol to guide and assist digital forensics professionals in investigating crimes committed on or via the deep and dark web, the protocol named D2WFP establishes a new sequential approach for performing investigative activities by observing the order of volatility and implementing a systemic approach covering all browsing related hives and artefacts which ultimately resulted into improving the accuracy and effectiveness. Rigorous quantitative and qualitative research has been conducted by assessing D2WFP following a scientifically-sound and comprehensive process in different scenarios and the obtained results show an apparent increase in the number of artefacts recovered when adopting D2WFP which outperform any current industry or opensource browsing forensics tools. The second contribution of D2WFP is the robust formulation of artefact correlation and cross-validation within D2WFP which enables digital forensics professionals to better document and structure their analysis of host-based deep and dark web browsing artefacts.

## 1. INTRODUCTION

With the ongoing development of the internet and the increased use of digital devices, crime has become a more digital phenomenon. Users with more sinister and unlawful intentions are accessing areas of the internet that are concealed from the public by being unindexed as illustrated in Figure 1. In contrast, everyday internet users access websites using a standard web browser, and the dark web uses numerous layers of encryption to encrypt all traffic, and services such as the TOR, FREENET, WATERFOX, and TAILS are used to access it [18]. As a result of these layers of encryption, the dark web offers users a high level of anonymity. This has resulted in several dark web marketplaces offering illegal goods like drugs, weapons, false passports, and more. In addition, users are further anonymized by cryptocurrency payment methods, such as Bitcoin, and encryption when browsing the dark web [13]. The term "dark web" is often used interchangeably with "deep web," "dark net," and "Invisible Internet Project". However, the "Dark Web" refers to websites hosted within overlay networks typically inaccessible without dedicated “Privacy and Anonymity” web browsers. Its most important feature is that service users remain anonymous; neither a website provider nor a visitor can identify the service provider. On the other hand, the deep web is any internet information or data that cannot be found using a search engine. In addition, some estimates say the deep web is much larger than the visible or surface web. The term "dark net" refers to the portion of the IP address space that is routable but not used [3-5]. The dark net is most associated with overlay networks that provide anonymous network connectivity and services. In addition, the Invisible Internet Project (I2P) is an unknown peer-to-peer network layer that uses layered encryption and garlic routing, a variant of onion routing, to ensure the anonymity of communications. The new generation of anonymity and privacy browsers, such as TOR and I2P, rely on a complex implementation of the onion routing topology initially introduced by the US Navy Research Lab in the mid-1990s to conceal the user's IP address [20]. As a result, attempts to trace or identify the user online by relying on traffic capture are nearly impossible. TOR and other networks are designed to protect against tracking, profiling and eavesdropping attacks, providing privacy and anonymity using mainly cryptography such as TOR multi-layer encryption in conjunction with an “onion” routing network deployed by tens of thousands of volunteer networks to direct traffic over the internet so a user identity can be kept hidden from network interceptors.

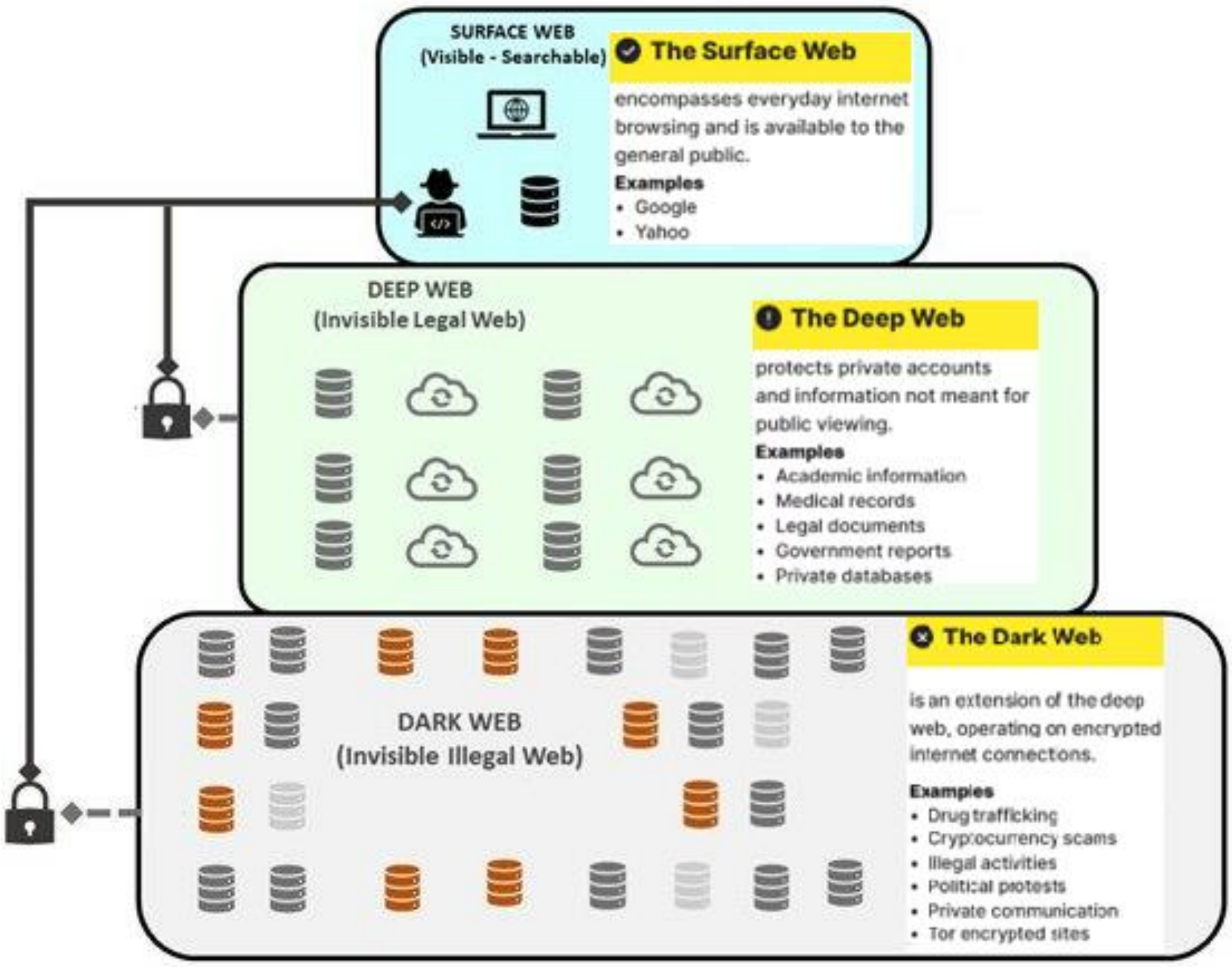

**Figure 1:** Anatomy of the Surface, Deep and Dark Web in the context of Visible and Invisible content.

**A. Research Context**

This study will briefly introduce the use of Privacy and Anonymity preservation internet browser such as TOR, FREENET, TAILS, WATERFOX to navigate the deep and dark web for illegal activities, this work has a special focus on TOR browser as one of the most used tools amongst criminals and thus will be used to simulate activities for research purposes to mimic the real-world situation. In addition, we will briefly describe the legitimate use of the same technology for security and privacy purposes which lead to the setting of a borderline between the two usages. Given the high volume and complexity of threats emanant from the Deep and Dark Web (DDW), we decided to explore existing methodologies used by law enforcement agents in investigating DDW crime activities. In addition to the comprehensive study, we will create different forensic scenarios to test the proposed deep and dark web forensics protocol. The internet is often described as consisting of three parts: the surface web, the deep web, and the dark web. While the terms "deep web" and "dark web" are regularly used interchangeably, it is helpful to highlight the complementary nature and not the interchangeability of these two terms.

**B. Research Scope**

Even though law enforcement and cyber security practitioners recognize that the deep and dark web doesn’t has a “monopoly” on online threats or negative effects/effects such as extremist views, criminal discussion, child pornography, terrorist propaganda, malicious hacking tutorials, stolen data marketing as hyped in media, it remains the most technically challenging online tracking activities for law-enforcement agencies across the world and highly popular amongst criminals [5]. This research intends to formally address the deep and dark web forensics process and proposes a robust protocol for forensically identifying, extracting and analysing deep and dark web criminal activities from users' machines and devices. It will investigate the digital forensics process when examining dark web content, methods of accessing it, types of illegal activities, and the use of anonymity and privacy for the benefit of cyber obfuscation on the dark web. Also, this research will critically examine the existing methodologies and methods currently adopted in dark web forensics. This research proposes a complete and comprehensive protocol, combining existing and new techniques and was designed, implemented and tested around a few hypotheses and scoping choices notably the use of TOR to simulate deep and dark web browsing activities and the testing limitation to the host (end-user) device investigation only. We will then test the proposed protocol in different scenarios, including various operating systems and internet browsers. As a result, a novel and comprehensive protocol for assisting digital forensics professionals in conducting deep and dark web artefacts forensic investigation from the host side was proposed. It is worth highlighting that even though the Deep and Dark Web is accessible using a regular internet browser such as Firefox, criminal activities are often carried out using a special browser such as TOR to guarantee privacy and anonymity which offer cyber criminal a sense of security and un-traceability from against law enforcement agencies, thus the investigation of artefacts lefts on the regular browser is outside the scope of this work.

**C. Paper Structure**

This paper is currently divided into five sections, each of which contains information about issues pertinent to dark web forensics. Section 1 provides an overview of the research, its scope and motivation. Section 2 presents in-depth background research on the subject, delves into the various deep and dark web forensics landscapes and discusses findings that, to some extent, contribute to this paper. Section 3 covers the adopted research methodology and justifies the choices made in this work. Section 4 summarises the novelty of the paper, notably the impact of D2WFP impact on current practice. Section 5 tackles the Order of volatility in Digital Forensics and Incident Response and its impact on web browsing investigation output. Section 6 introduces the proposed high-level protocol that mimics the four standard DFIR phases. Section 7 describes the detailed D2WF Protocol and provides backing evidence for the Order of tasks and sub-tasks. Section 8 explains the designed forensics scenario and the research input in the form of Datasets. Section 9 describes the testing phase in terms of forensic processing, examination, and analysis. Section 10 presents the results with some criticalities. Finally, section 11 concludes the research with a final reflection on how the research could be improved.

## 2. LITERATURE REVIEW AND RESEARCH GAPS

The literature review dives into the background and technical details of the deep and dark web, describing its development as a tool for new types of criminals to grow their activities. The information was gathered from several reliable sources, including books, journal papers, reports, conference proceedings, and web pages.

**A. Related works**

The internet is a crucial tool for facilitating modern societal life. Through primary, traditional web browsers and an internet connection, users can access websites for social networking, online shopping, video streaming, public news, research, and more via the "surface web" [19]. The crucial distinction between the "surface web" and the "clear web" is that the latter phrase refers to the portion of the internet that can be indexed by any traditional search engine [2]. The deep web stands in opposition to the surface web. The deep web is a collection of websites, communities, networks, and intranets that are purposefully not accessible by standard online browsing, in contrast to the surface web, which makes its material searchable by ordinary search engines [3]. However, because of its anonymity, the dark web offers services to those who could have illegal intentions over a global network that is essentially untraceable [19]. This raises an ethical problem due to the growth of the dark web, which has given rise to lucrative sectors like malware services, child pornography, and illicit internet markets. In addition, the most prominent entrance to the dark web is TOR [8]. I2P is one solution used to access the dark web. Still, other anonymous software programmers like Freenet and OpenBazaar allow anonymous communications by storing and exchanging data via the machines linked to the network rather than utilising centralised servers [7]. Terrorist groups use the dark web and the Islamic State to propagate propaganda, attract new members, and share data and films that support terrorism [9]. The dissemination of child pornography and unlawful online markets for illicit goods (drugs, weapons, passports) are the two crimes that are most frequently committed through the use of TOR [5]. However, other crimes include renting assassins and trafficking humans. One of the most well-known instances was the 2011 emergence of the drug bazaar on the dark web called "Silk Road." The FBI shut it down in October 2013 after it had operated for two and a half years and processed over $1,200,000,000 in transactions [8]. The fact that criminals frequently feel safer transacting on the dark web than on the street is one of the significant elements causing this crime as has been pinpointed in two recent studies on policing the dark web [8-9]. Even if there is a higher likelihood that law enforcement will halt the business and take it offline, additional sites will replace it [10]. Police and Law Enforcement agencies shutdowns are one of the reasons why many dark websites are only active for a limited time, usually between 200 and 300 days, according to analysis [12]. Other elements that have influenced the rising use of online criminal marketplaces include the growth and acceptance of cryptocurrencies [16].

A 2016 study by RAND estimated that the three biggest online criminal markets at the time accounted for 65% of all crypto market listings. Cryptocurrencies like Bitcoin and Ethereum allow users to make payments anonymously as an additional layer of security [10]. In [11], a study was conducted to gain a deeper understanding of the dark web and its impact on people's lives. The researcher achieved the aim of his paper through the provision of various methods of access, the listing and description of available websites, and the provision of a list of precautions people should take before surfing the dark web. The article also discusses illegal activities and crimes committed on the dark web, its ethical and unethical sides, its pros and cons, and how legislative agencies and security agencies can administer the dark net to secure society. The research produced a robust research model on the dark web because it highlights methods of accessing the darknet through various specific configurations, software, and authorisation. Also, the research considered the precautions a person should take while attempting to delve into the darknet world, especially for a first-time user, and the associated risks, such as malware and the loss of personal data and identity [20]. Finally, the researchers discuss the various applications of the dark web, emphasising its advantages and disadvantages but failed to address the security and enforcement part of the Deep and Dark web thus finding were more about precautions part which they should have considered illegitimate use of privacy-preserving browsers and the privacy and security offered to criminals because of their non-malicious activities. Therefore, this research focuses primarily on platform tools such as TOR, FREENET, TAILS, WATERFOX and other methods put in place to enhance dark web security, such as ISP Invisible Internet Project (ISP), Enhanced Firefox with the usage of HTTPS Everywhere and VPN [5].

In [4], the authors examine and track the nature of activities on the dark web using digital forensic (DF) tools, particularly in light of the evolving technology that renders traditional tools obsolete. They achieved their research aims by executing sampled, predetermined sites on a closed-paraben Electronic Evidence Examiner (E3) using a TOR browser on an Ulefone Note 7 mobile device. Also, they assessed and analysed the software's ability to track benign content. According to [4], in digital forensics, tools help to crawl and sift through large volumes of data that users generate. These tools, including The Sleuth Kit (TSK) and E3, are more accurate and save time since they crawl different types of data that are voluminous [15]. The research considered downloading benign content from sites that would encourage criminal activity and the ability of the tools to gauge illicit activity by using legal models as mirrors. However, the investigators ignored the use of various software, such as Encase and Forensic Toolkit, different browsers, various mobile devices, and devices running iOS or Windows OS, which could have yielded a different result. The research considered timestamps for the TOR browser and the usage of the application on the device while using E3 software. However, this research teaches us the need to use broad digital forensics methods for better results.

In [17], the researchers conducted a systematic literature review (SLR) on the dark web, its crimes, and better methods for control, as well as how investigators can leverage its main feature (anonymity) for crime control. The SLR approach helps them achieve their goal for this paper using a definition of the research questions, a systematic review of the literature, a search for relevant data sources, the extraction of data, and analysis for meaningful reporting. The researchers state that such a study helps provide knowledge about web crime spikes but also helps investigate the dark web's impact on users' lives [20]. Moreover, the study helps evaluate the challenges raised by methods used to control crime and their weaknesses. The research model was good because different techniques could be garnered from it, including a statement of the hypothesis, determining the selection of materials, and analysing and synthesising data. The researchers classified data from 69 papers into two categories that answered their hypothetical questions: an outline of threats from illegal activities, methods of locating criminals on the dark web, and steps to control crime. Nevertheless, the evolving trends, such as the use of newer technology in the face of technological advancement, have been overlooked. In essence, this paper gives the reader an in-depth knowledge of the crimes on the dark web.

In [21], the researchers used a honeypot, a protective tool, to investigate and gather data from malicious actions on the dark web. The goal of this research is to use and monitor two different honeypots; they research and produce the honeypots on the dark web over seven months, then analyse the information gathered on cybercrimes and detect the prowess of the protection tool. The structure of this research allows a reader to see different applied methodologies, such as testing prevention tools, which in this case were two; background information about the dark web and the types of honeypots; an overview of related works; and analysis of the data collected from the tools. The research considers using virtual machines (VMs) that contain the ChatRoom web server, web-based honey, and ELK log server, ensuring security and flexibility. The researchers also confirmed that the VMs were secure from private escalation with clean snapshots, provided that the VMs reverted to daily for security purposes.

Additionally, the researchers ensured that log servers do not go off while they are set on a secured VM that guarantees safety [14]. However, to avoid fake logins and script attacks, the researchers used a captcha to ensure user authentication to the chatroom. However, future research must avoid improper comment data filtration because it leads to remote code implementation through unauthorised default settings. This study helps one to learn the use of different honeypots for the best results and understanding of dark web crimes and cyber-attacks [22]. In [5], the researchers examined the importance of memory forensics, also known as a forensic study of the computer dump, a widely accepted part of the incident report process for investigation. They achieved the goals of this research through a discussion of recent trends and problems in online crime, an outline of the technical background and discussion of the expansion and growth of memory forensic techniques and tools, and a review of the current memory forensic scheme and its drawbacks. They opine that memory forensics tools for analysing memory are essential because they help identify specific areas that have been compromised by malware or help uncover users' digital footprints. They also direct analysts to areas to focus on during malware investigations. According to [5], if an investigator does not know where to look, the investigator may waste valuable time as the crawler sifts through large data sets. The advancement of capacity and media poses the problems of accessing data and its analysis and the development of frameworks without forensic memory capabilities, such as the OSX system in Apple products, which are attractive to criminals. This paper sets an excellent example for anyone interested in helping forensic investigators with memories. They provided discussion, challenges, and a technical overview of the study and looked into emerging trends that are important for shaping future solutions in digital space. As a result, this study demonstrates that a successful exploration of digital scenarios emphasises the importance of memory forensics, and an investigation into specific applications that aid in understanding their actions is unquestionably an advancement for that field.

**B. The Landscape of Dark and Deep Web Forensics**

Frameworks and techniques for dark web forensics are scarce compared to other types of computer forensics (based on what is in the public domain). However, some have made an effort to map this out. According to [13] and [15] research journals, browser forensics examinations focus on performing more conventional operations on the computer, such as database, RAM, network forensics, and registry. In contrast, dark web forensics should consolidate two fundamental areas: forensics connected to TOR activities and forensics linked to Bitcoin transactions [17]. The focus of Bitcoin forensics, however, is on monitoring payments [24-26]. The techniques suggested are illustrated in the figure below. The data mentioned above offers a starting point, but there is a general absence of knowledge and a reliance on using subpar open-source technologies.

| Browsing Artefacts | Evidence Location | | |
|---|---|---|---|
| | File System | RAM | User and System Configuration |
| URLs | No | Yes | Yes |
| Website Content | No | Yes | No |
| Search Queries | No | Yes | Yes |
| Bookmarks | Yes | Yes | Yes |
| Cookies | No | No | No |
| Email Addresses | No | Yes | No |
| Email Content | No | Yes | No |
| Usernames | No | Yes | No |
| Passwords | No | Yes | No |
| Download Files | Yes | Yes | No |
| Usage/Session | No | Yes | Yes |
| Timestamps | Yes | No | Yes |

Figure 2: Summary of Internet activities Artefact's location per category in Windows and Linux distributions.

This research direction will be significantly aided by studies and experiments carried out with better tools and in greater depth. This was accomplished through a research study from Marshall University that expanded on some of the earlier concepts, introduced the AccessData Forensic Toolkit as a tool for analysis, and went further in-depth on the testing environments employed [6]. The use of RAM forensics by getting a live memory dump of a suspect's computer (or, in this case, a virtual machine created by using an FTK imager) and then analysing this through FTK yields promising results, with index searches relating to TOR yielding results in both the memory dump and page file areas [6]. The study also considers the idea of observing registry and cache changes and contrasting a virtual machine before, during, and after using TOR, which did produce some information allowing an investigator to suspect TOR is used (although nothing concrete), as well as using Wireshark to compare how the traffic compares between a standard browser and TOR, with there being a discernible difference (a comparison of the protocol hierarchy showed a clear distinction) [6]. These techniques aid in shaping what should be contained in the D2WFP and are helpful for later testing in this research. These forensic procedures are only sound if the device is captured while still switched on as TOR browser processes are allocated in the live memory, and only live RAM capture can preserve such crucial artefacts.

In [1], similar research was made to show off forensic methods for TOR browsing on Windows 10 and Android 11 devices. The procedures are carried out following the earlier reports, with an emphasis on the registry, memory, and local file system in line with the idea of a scenario set up on a virtual machine. When thinking about the methods to be added to the framework, this offers a sound basis for what has to be tested. In contrast to the previous report, which was not simple, the entire report has numerous figures and tables that demonstrate the tests being conducted. The most helpful report was found as a result of this; it was a doctoral dissertation submitted to Dakota State University which focused on finding any Dark Web artefacts on a machine and creating a framework around this [15]. The depth of information in the report extends beyond the scope of this one. Still, it provides a solid foundation for what needs to be taken into account and lists every possible place where TOR-related activity might be hiding on a device. In [13], the authors focused on internet browsing artefacts extraction during live memory forensics and noted that EnCase and FTK are proven to be reliable in terms of covering, exploring, and analyzing live memory artefacts compared to other publicly available tools but remain inefficient in context of dark web browsing when TOR is used and impacting artefacts collection and interpretation notably by generating a large amount of unreadable data. The new generation of privacy-preserving browsers such as Brave and TOR are even more challenging when it comes to browsing data forensics examination as the amount of encryption and encapsulation reduces the amount of readable data and this limit the investigation as illustrated in the cryptographic upgrade of TOR which took effect early 2017 [27].

### C. Internet Browsing Forensics Evidence, Techniques and Tools

As the background study has indicated, the issue of dark web forensics is primarily related to the process of investigating the digital equipment used to carry out criminal activities. The issue is gaining importance along with the growing internet privacy and security landscape. The dilemma is the dual use of the technology and tools to access and maintain deep and dark web content, such as TOR and I2P. Furthermore, the current methods used fall short of effectively investigating DDW cases and thus combating such crime. The Order of volatility plays a central part in this field. The RAM and Caches on the running evidence machines should be systematically extracted using the AccessData FTK imager or other equivalent tools. On the other hand, the volatility application is used to analyse the RAM further to identify the types of applications running on the process ID, downloaded documents, and visited websites. FTK Registry Viewer and Registry Editor are used on the host machine to analyse evidence of TOR installation, the last executed date, and other attributes that might be of significant value to investigators. Wireshark and Network Miner are tools for extracting data from networks. However, Wireshark lets users view packets of data as they travel through a computer network. As a result, the PCAP files enable investigators to gather and analyse web traffic information and network connections. The Internet Evidence Finder searches the SQLite Database for evidence related to users' visits to web content and identifies TOR browsing histories. The FTK and UFED discovered the application-related data, including usage session, timestamp, and cookies. To sum up, we elaborated a summary of the evidence identification, technique and tools used in investigating deep and dark web. Table 1 illustrates the adopted Deep and Dark web forensics evidence, techniques, and tools.

Table 1: A summary of Deep and Dark web forensics evidence, techniques, and tools.

| Evidence | Techniques | Tools | Purpose |
|---|---|---|---|
| **Host Machine** | **Live RAM** | Exterro<br>FTK Imager<br>Volatility Framework<br>Magnet AXIOM | Obtain the description of the types of URLs, wikis, and visited deep web websites, as well as other downloaded content. |
| **Host Machine** | **File System Forensics** | Exterro<br>FTK (Registry Viewer)<br>Windows Registry Editor | Regshot and analysis to obtain evidence of TOR installation and last executed date and other attributes |
| **Network Traffic** | **Network Forensics** | Wireshark<br>Network Miner<br>Kroll KAPE<br>Cellebrite UFED | Gather and analyse evidence of web traffic, established VPN and proxy connections and information on network connections available |
| **Host Machine** | **Browser Forensics** | Magnet Internet Evidence Finder (IEF)<br>Dumpzilla | Locate, extract and retrieve evidence related to users or visited dark web content and activities |
| **Application Forensics** | **Applications and Transactions** | Exterro FTK<br>Cellebrite UFED<br>Magnet AXIOM | Recover applications' related data, including usage session, timestamp and cookies |

## 3. RESEARCH METHODOLOGY

The first step in this methodology is to emulate the forensic scenarios carried out at the London Metropolitan University Cyber Security Research Centre. We run the scenarios on various operating systems and web browsers. The chosen devices and machines for the scenarios are a Windows 10 machine, Kali 2021 version 3, Android 11, and iOS 14.2. For forensics imaging, analysis, and examination, this research uses fully licensed Access Data FTK, Magnet AXIOM, and Cellebrite UFED tools. TOR is installed on all devices to access the dark web, store login information, and browse the web with Chrome and Microsoft Edge.

A forensic image is created due to the massive data generated during dark web browsing activities, which will then be analysed using FTK and Magnet. After the forensic images are created, an anti-forensic tool called BleachBit is installed on all devices and used to clear any browsing artefacts before creating new forensic images. Subsequently, we generate another browsing history on all devices before creating and storing away another set of forensic images. Fully-licensed FTK and Magnet are used to analyse the forensic images for Windows 10 and Kali 2021.3. Simultaneously, forensic images for Android 11 and iOS 14.2 are analysed using the Cellebrite Physical Analyser software packages following the principles outlined in our proposed protocol. We then quantitatively analyse the data collected from the forensic scenario analysis and examinations. The outcomes are tabulated to compare the standard automated process to our proposed protocol. This research aims to compare Deep and Dark Web Forensic Protocol Results to Regular Automation after Anti-Forensic Bleach-Bit using FTK for Windows 10, AXIOM for Kali 2021.3, and UFED for Android 11 and iOS 14.2. The results obtained from the analysis and examination show an apparent increase in the number of artefacts recovered when adopting our proposed protocol compared with regular automation and framework-based automated investigations. The desired outcome of this research is the development of a solid protocol for dark and deep web forensic investigations that are clear, efficient, and effective enough to be used as a reference by cyber forensic investigators. The protocol will aid forensic examiners and, more importantly, analysts in their work by relying on current industrial (licensable and non-licensable) tools and frameworks to which they have access.

## 4. **RESEARCH CONTRIBUTIONS AND NOVELTY**

The first contribution of this research is to examine and assess the efficacy of our proposed protocol for assisting and improving dark and deep web forensics investigations. The findings and results show that the proposed protocol extracts more artefacts than standard automation. The second contribution is to analyse and document how to investigate dark and deep web forensics using the strategy outlined in our proposed protocol, such as extracting artefacts from operating systems, network traffic, malicious browser plugin behaviour, and memory dump analysis. The third contribution of this research is to draw attention to the flaws of privacy tools, particularly TOR, such as its tendency to take longer to load when connecting to available node servers, slow performance because of routing data through different nodes, and the fact that data inserted on web pages are not encrypted. Lastly, the fourth contribution is significant because we value knowledge exchange and aligning teaching to industrial practice, which will be a direct output of this research. We aim to use both the proposed protocol D2WFP and evidence files created from simulated deep and dark forensics scenarios in the new cyber threat intelligence and dark web forensic curriculum to be delivered for undergraduate, postgraduate and CPD students as well as for conducting future research works.

The initial protocol is elaborated based on theoretical knowledge and has been inspired by previous research and established forensic principles, as explained in Figure 3. We then proceeded with the elaboration of eight (08) forensic scenarios created on different OS, respectively Windows, Linux, Android, and iOS, where we used TOR to access emulated deep and dark web addresses and content. At the end of the simulation, we dealt with forensic imaging by running an anti-forensics tool to delete artefacts and forensic re-imaging using FTK imager and UFED Extractor.

| Process | Description |
| --- | --- |
| **Identification** | Distinguish between valuable and worthless evidence and primary and secondary memory evidence, such as additional MicroSD or cloud. |
| **Preservation** | Evidence is cloned while write-protected and stored with security. At the same time, the forensics processing will apply to the clones or images (e.g., a UFD image for mobile evidence and an AD1 image for computer-based HDD evidence). |
| **Analysis** | Examining and analysing data by applying forensic techniques and solutions to the forensic copy of the evidence in addition to data identification and indexation. This includes deleted data recovery, raw data carving, and partial data reconstruction. |
| **Presentation** | Reporting must effectively communicate all pertinent information in a way that non-experts may understand. |

Figure 3: Standard Digital Forensics Process

We then moved to the forensics examination, recovery and analysis using FTK, AXIOM, and UFED Physical Analyzer (PA) software packages following the application of the principles outlined within the protocol. The identification phase differentiates between valuable data such as transaction records, timestamps, login credentials, browsing caches holding posts and/or comments as well as networking activities including pertinent IP addresses which are often related to higher layers in the Order of Volatility (OoV) and primary and secondary memory evidence, such as additional MicroSD or cloud storage. The evidence is cloned, preserved while write-protected, and securely stored when forensic processing is applied to the clones. We reviewed and analysed the data using forensic methods and solutions to the forensic copy of the evidence. However, the reporting must properly communicate all pertinent information in a way that non-experts may understand.

## 5. ORDER OF VOLATILITY IN CYBER FORENSICS

We needed to grasp and understand the existing forensic processes and procedures in their parts covering web browsing artefacts in general before elaborating on the investigative protocol aimed at the realm of dark web forensics. The phases of the computer forensics procedure and the OoV are two essential elements that recur regularly. These are crucial factors to take into account when creating any framework for digital forensics since it dictates how investigators should handle a case from start to finish. The OoV defines which data should be collected first because the nature of the data makes it quickly adaptable through simple machine activities.

| OoV | Explanation |
|---|---|
| 1 | Cache and Registers |
| 2 | Routing tables |
| 3 | ARP cache |
| 4 | Process table |
| 5 | Kernel statistics and modules |
| 6 | Main memory (RAM) |
| 7 | Temporary file system |
| 8 | Secondary memory |
| 9 | Router configuration |
| 10 | Network topology |

Figure 4: Order of volatility in Digital Forensics and Incident Response.

Although there is no acknowledged volatility hierarchy, Figure 4 illustrates the Order of Volatility in browsing forensics and covers the specific example of Windows operating systems. This later indicates that examples of data other than RAM should be considered when gathering evidence. Investigators must take the computer forensics procedure into account. Although there are no set standards for this, the outcomes of all reported approaches are comparable.

## 6. THE PROPOSED HIGH-LEVEL PROTOCOL

With these fundamental computer forensic principles identified, we can establish the protocol. A universal protocol is developed for the scope and objective of this research so that researchers can use it with various operating systems and software programs. In-depth needs for scenarios like various operating systems (Windows, Mac, Linux, Live OS) or software packages for dark web access can be determined through further research on this topic (I2P, Freenet). The suggested protocol, known as the D2WFP, can be found in Figure 5. This protocol will serve as an essential manual for people who seek to examine possible evidence linked to crimes committed on the dark web, stressing the kinds of forensics that investigators should perform and the proper sequence in which to do so. The crucial procedures for locating the evidence and protecting it have been emphasised. If this is done incorrectly, the forensic analysis may produce tainted evidence that is not admissible in court. Figure 5 illustrates the alignment of the proposed dark and deep web forensics protocol with the standard digital forensics phase divided into four branches: evidence identification, acquisition and preservation, examination and analysis, and finding presentation.

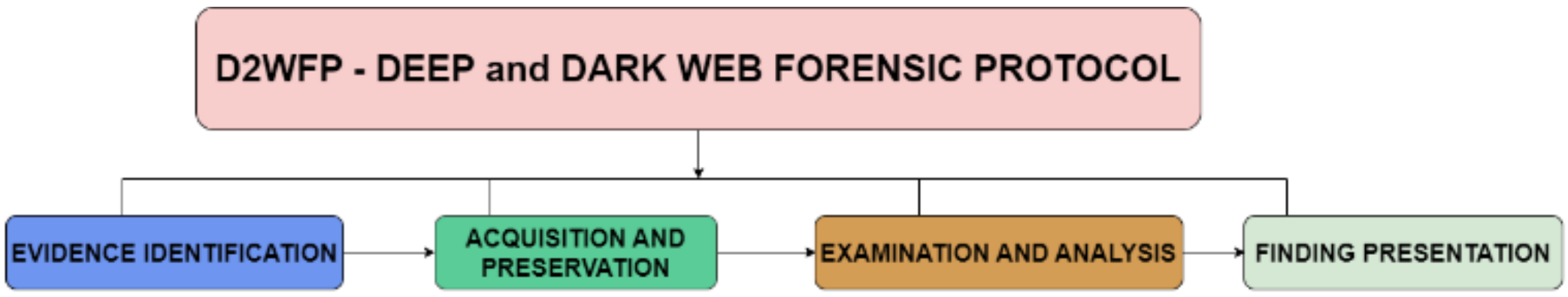

Figure 5: D2WFP - Deep and Dark Web Forensic Protocol Branches.

## 7. DETAILED D2WF PROTOCOL

### A. Evidence Identification

Evidence identification is vital to a trial in the legal system. As a result, identifying what makes a piece of evidence vital to building a solid case is essential.

- Identify and Collect Suspect Devices: Identifying potential evidence that may have been used in a crime is crucial to ensuring it will be admissible in court. This can be achieved by examining the crime scene, speaking to witnesses, or reviewing surveillance footage.
- Identify Extra Storages (e.g. Cloud and IoT): Once potential evidence has been identified, collecting and documenting accurately is crucial.
- Identify Extra-Storages (e.g. Cloud and IoT): During the evidence identification phase, it is critical to identify extra-Storage (e.g., cloud and IoT) that may contain vital data, and a process technique to extract this data should be developed.
- Determine the Required Hardware/Software: determining what is necessary to complete the evidence identification will include hardware, software, and storage requirements. Each task is unique, and it is critical to tailor the needs of the task to the hardware, software, and storage requirements.

**B. Acquisition and Preservation**

Acquisition is the process of gathering and collecting evidence, whereas preservation is the act of keeping, protecting, or preventing evidence from deteriorating.

- Establish a Chain of Custody: a "chain of custody" is a paper trail or chronological documentation that shows the seizure, control, transfer, analysis, and disposition of physical or electronic evidence. When gathering evidence, it is critical to maintain a chain of custody to ensure its integrity and admissibility in court.
- Acquire Volatile Data Following the Order of Volatility (OOV): Volatile data is information stored in a computer's temporary memory, and it is lost when the system is powered down. It is critical to acquire evidence in the Order of volatility to ensure that as much data as possible is recovered from a system. This means that the most volatile data, i.e. data in a system's temporary memory, should be acquired first.
- Acquire Physical or Logical File System Image: a physical file system image is a bit-for-bit copy of a storage device, such as a hard drive, that can be used to forensically examine the device's contents, whereas a logical file system image is a copy of data on a storage device that can be created without creating an exact bit-for-bit copy of the device. Logical file system images typically include only the files and folders required for a specific investigation.
- Import/Backup Cloud-Based Content: acquiring and backing up cloud-based content ensures that the evidence is not lost or tampered with if the cloud-based service is shut down or deleted.
- Extract Networking and Logging Data: investigators will often need to collect networking and logging data from reconstructing events or tracking down suspects. This information can be obtained by obtaining it from network devices or servers.
- Secure Evidence and Store Original Devices in Safe: Once the evidence has been collected, it must be secured and stored in a secure location. This will prevent the evidence from being lost or tampered with. Keeping the original devices in a safe location is also critical to ensure their integrity.

**C. Examination and analysis**

Examination and analysis are the reconstruction and interpretation of evidence and follow the tasks below.

- Analyse RAM and Examine URLs, Web Content, Search Queries, Logging Details, and Downloaded Files: Examine the running processes, visited websites, downloaded files, system logs, open files, and network connections to reconstruct events.
- Analyse Browsing Data, File System/Registry, Examine URLs, Bookmarks, Usage Sessions, Timestamps, and Caches: we can begin by looking at web browser data to get an idea of what the user is doing on the system. This provides an overview of the websites the user visits and the types of activities carried out. The -r option returns a grep list of all bookmarks saved by the user. We can use the grep -c option to print the number of times each URL has been visited. To see a list of all the cookies stored on the system, use the -f option with grep. We can use the grep -s option to see how many times each cookie has been accessed.

- Analyse Logs and Networking Cookies, Examine Downloads and Browsing: Examine the log files with the command-line tool grep to search for specific keywords in the log files. We can use the grep -r option to see a list of all the IP addresses that the user has accessed. We can use the grep -c option to see how many times each IP address has been accessed.
- Analyse Cloud Backup and Examine Logging, Browsing, and Search Queries: this search can reveal user activities during the Examination and Analysis phase. However, the information that is stored in the remote location through a network such as the Internet is further analysed, the logging details of the activities and who was involved are examined, and the search behaviour of the user is evaluated so that the investigator can organise and get more insight on the user's activities and the incidents that happened.
- Correlate Findings and Establish the Final Timeline: We reviewed the findings and looked for patterns or themes. If there are any gaps in the data, we conduct additional research to correlate the findings. Once we have a clear picture of what occurred, we create a detailed timeline of events that will serve as the foundation for the final report.

**D. Findings Presentation and Reporting**

The layout and reporting is the protocol's final phase, in which we summarise and draw a conclusion based on the evidence. Request Further Data (ISP, Web Servers): If Applicable, we can request more information from the ISP or Web Servers.

- Process Finding, Correlate Finding, and Remove Duplication: At this stage, we process all of the evidence files' findings and correlate them to remove any duplication.
- Summarise Findings in accordance with the Relevance and Acceptability to reconstruct the crime environment and ensure that the evidence is admissible.

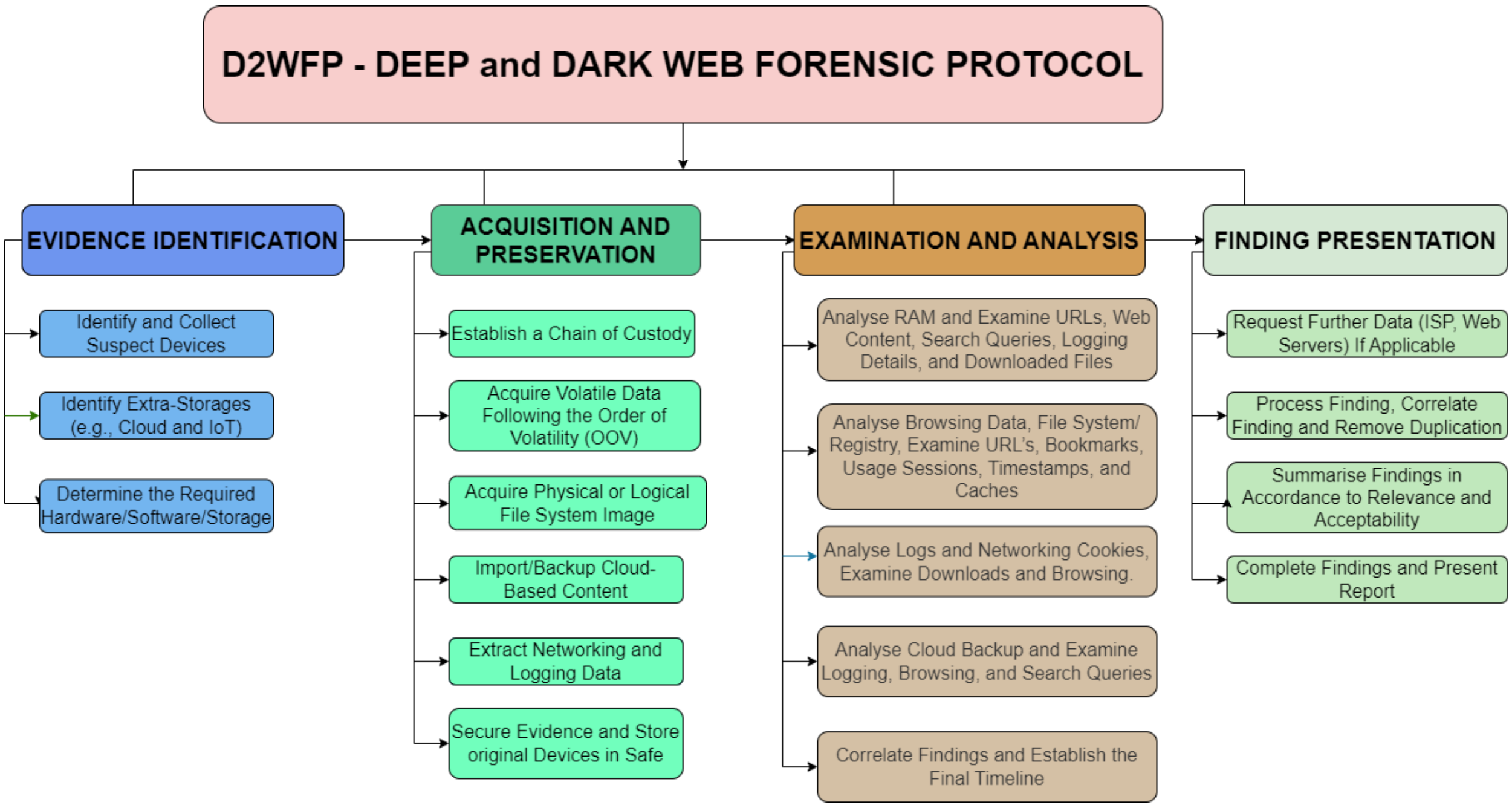

Figure 6: Detailled Proposed Protocol - Deep and Dark Web Forensic Investigation Protocol

- Complete Findings and Present a Technical Report: When all the above digital forensics processes have been completed, the findings need to be presented and put in an orderly document that validates the evidence collected, tracked, and analysed. The findings should be protected and kept in a safe place for access when needed. Security of such information is vital to avoid tampering or loss of data.

The overall proposed Deep and Dark Web Forensic Protocol (D2WFP) is represented in Figure 6.

## 8. FORENSIC SCENARIOS DESIGN AND DATASETS GENERATION

In this research, we aim to cover as many digital pieces of evidence as possible. We used the resources available to us to create realistic digital forensics scenarios by using four different types of devices, each running a different operating system and all using the TOR browser or an equivalent browser for Apple iOS. The first machine is running Windows 10, the second is running Kali Linux 2021, the third is a mobile device running Android 11, and the last is an iPhone running iOS 14.2. We asked certified law enforcement agents (LEA) with a track record to emulate cyber-criminal activities at the local network by using a basic client-server setting where replication of previously reported and shutdown hidden wikis are hosted in the server and accessed by users using TOR installed in the four machines. For consistency purposes, LEAs were asked to execute the same scenario in terms of browsing, logging, saving credentials and creating bookmarks. All the following predefined scenario LEAs include a local replication of shut down hidden wiki as illustrated in Figure 7. All these activities were carried out on simulation equivalent setup for teaching and research purposes, and LEAs adhered to strict research guidelines to avoid any misconduct. After completing the forensics case, we moved on to forensic evidence identification and acquisition, mainly by performing live memory capture (RAM) with the FTK Imager, UFED, and capturing networking data from the access points. In addition, we used the FTK Imager to acquire forensic bit-stream HDD images for the two hard drives used in the Windows 10 and Kali Linux 2021 machines.

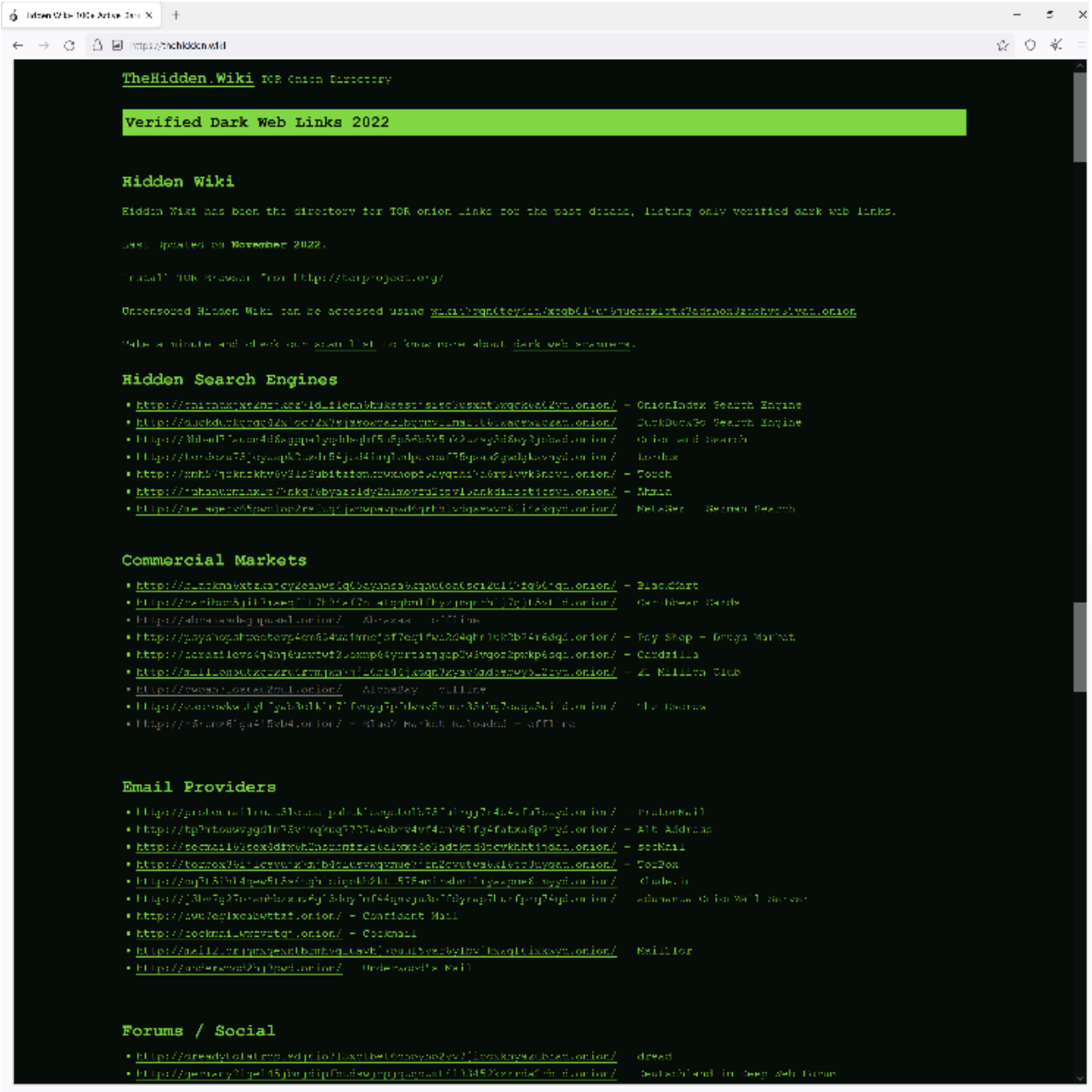

Figure 7: Deep and Dark web browsing emulation from The Hidden Wiki directory using the TOR browser.

We did, however, perform a logical plus file system extraction for Android 10 and iOS 14.2 devices because physical extraction was impossible as neither device was jailbroken or rooted. Following the completion of the first round of data acquisition, the next step was to apply anti-forensics activities through the use of BleachBit to remove any existing browsing artefacts. Imaging of the device was performed using the UFED Touch 2 tablet provided by Cellebrite and is demonstrated for mobile devices and OpenText TX1 for computer hard drives, as illustrated in Figure 8.

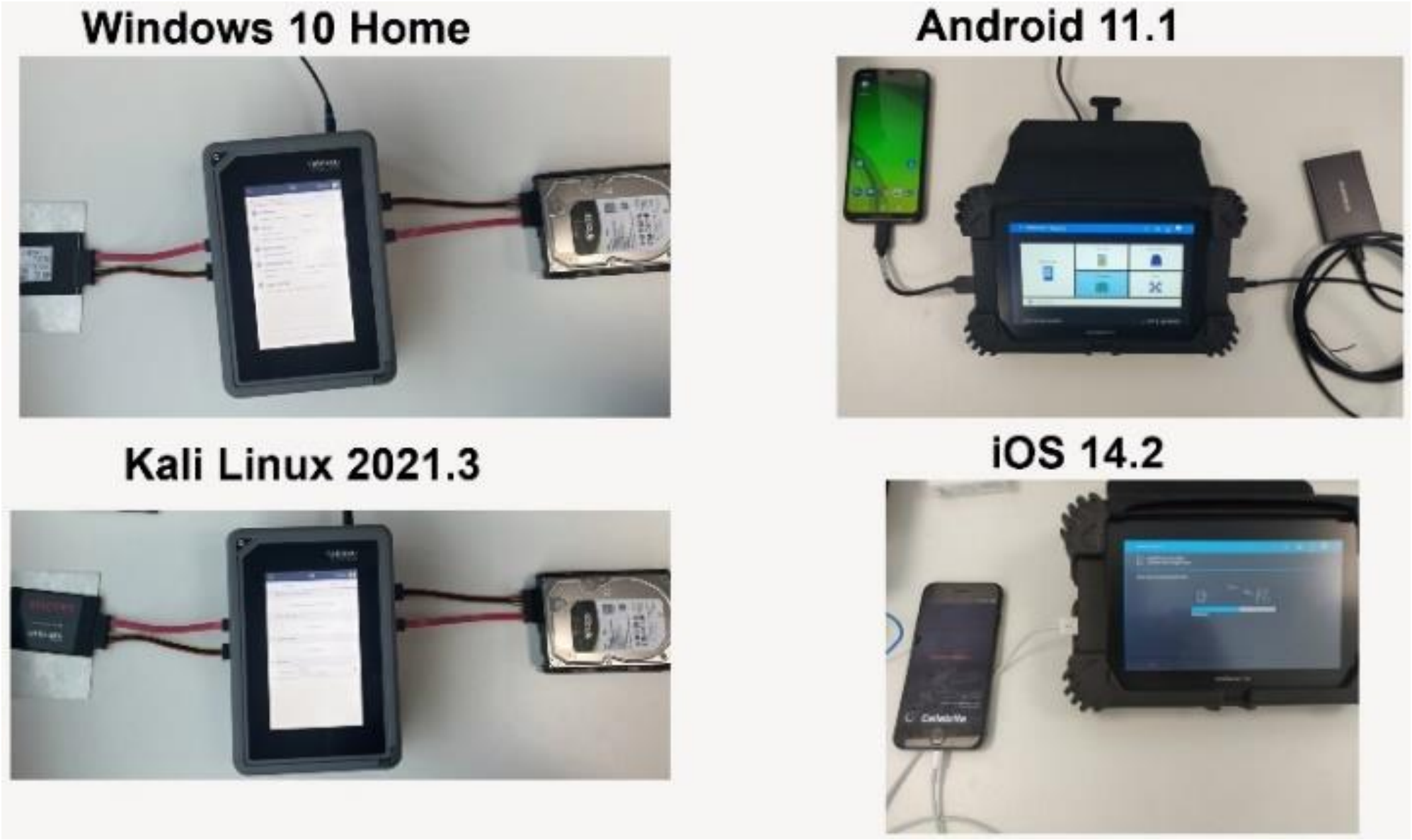

Figure 8: Microsoft Windows 10, Linux Kali 2021.3, Apple iOS 14.2 and Google Android 11.1 forensics imaging.

We performed anti-forensics tasks by installing BleachBit, a browsing anti-forensics software, on all four devices before acquiring the image files again. We performed several rounds of deletion of internet browsing activities to determine whether our proposed protocol is applicable in anti-forensics cases. We then jailbroke the iOS device and rooted the Android device before performing a second round of data acquisition on the mobile devices. As a result, we created a new set of images for the hard drive running Windows 10 and Kali Linux 2021, as well as complete physical extraction (available after jailbreaking and rooting) for iOS and Android smartphones. Figure 9 illustrates the use of BleachBit to delete caches, temporary files and browsing-related artefacts.

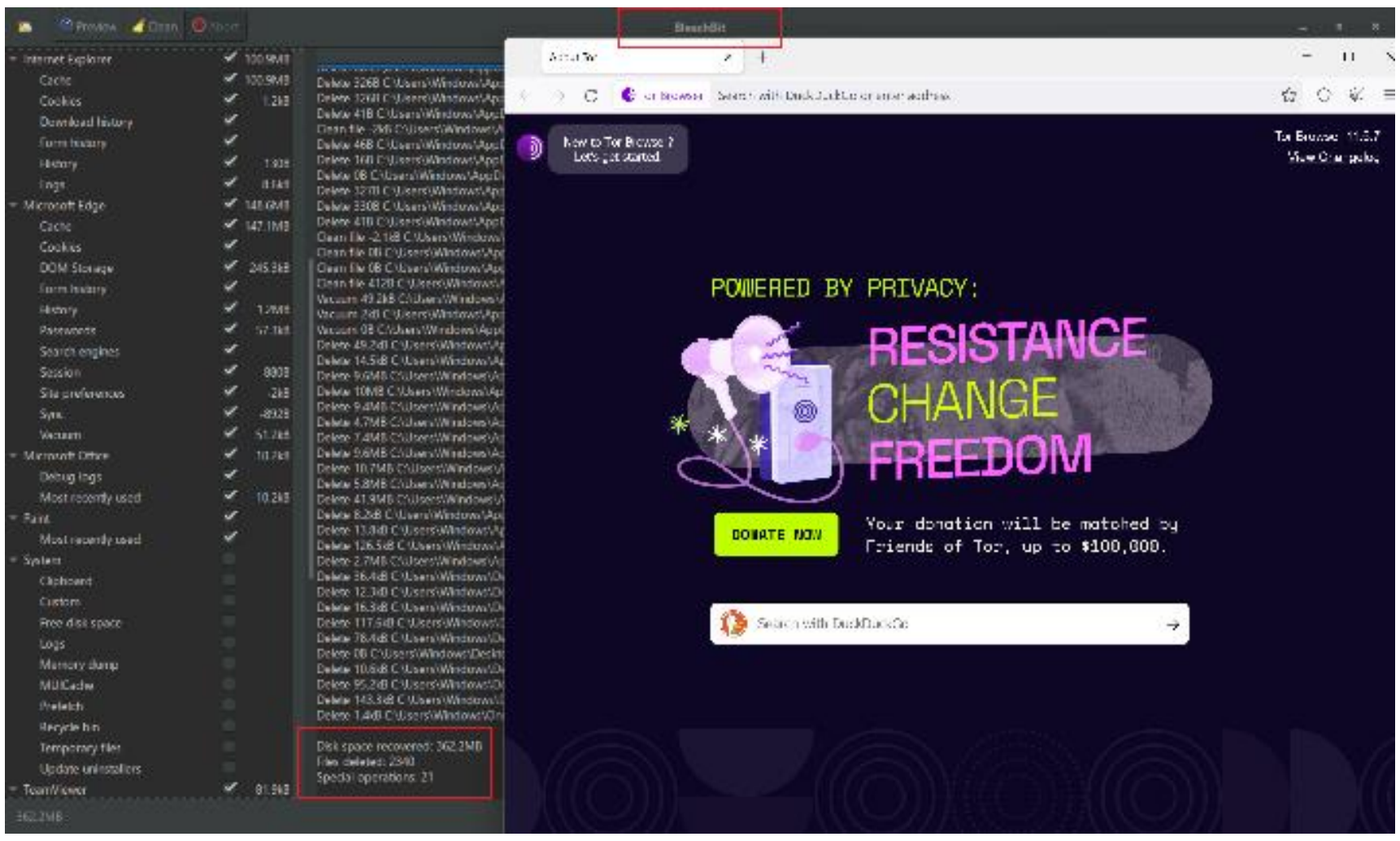

Figure 9: Applying Anti-forensics by deleting artefacts using the Bleach-Bit tool.

We then analysed and compared artefacts such as browsing history, security and logins, cache and temporary files, SQLite DB forms, and downloads that we could determine from all the created forensic images. Following image analysis, the primary quantitative data analysis technique is hypothesis analysis, which compares the study hypothesis with the findings using sample data from the developed images. However, the quantitative analysis generates tables highlighting the artefacts discovered for each image, such as cookies, login information, and browsing history. We subsequently investigated the testing results by confirming the extent and amount of information retrieved about TOR activities. The results are presented visually in Tables 2 and 3. We highlight the number of recoverable artefacts for various types of devices and the impact of the tools used in removing the web-based artefacts.

Table 2: D2WFP - Deep and Dark Web Forensic Protocol Result compared with regular automation using FTK for Windows 10, AXIOM for Kali 2021.3 and UFED for ANDROID 11 and IOS 14.2.

| ARTEFACTS | Protocol | WINDOWS 10 | LINUX KALI 2021 | ANDROID 11 | APPLE IOS 14.2 |
|---|---|---|---|---|---|
| BROWSING HISTORY | Regular | 1094 | 1086 | 1238 | 991 |
| | D2WFP | **1325** | **1631** | **1562** | **1133** |
| SECURITY & LOGINS | Regular | 30 | 30 | 30 | 30 |
| | D2WFP | **53** | **77** | **66** | **34** |
| CACHE & TEMP | Regular | 6732 | 5328 | 3627 | 4381 |
| | D2WFP | **10938** | **9677** | **7201** | **6320** |
| SQLITE DB FORM | Regular | 227 | 328 | 196 | 188 |
| | D2WFP | **636** | **801** | **786** | **295** |
| DOWNLOADS | Regular | 106 | 106 | 112 | 112 |
| | D2WFP | **317** | **429** | **299** | **243** |

* Results are generated automatically without accounting for empty or invalidated entries

Table 2 illustrates the number of artefacts by category obtained using DFIR tools and D2WFP before any anti-forensics activities have taken place. On the other hand, Table 3 illustrates the number of artefacts by category obtained using DFIR tools and D2WFP after applying anti-forensics measures by deleting the internet and browsing artefacts using Bleach-Bit.

Table 3: D2WFP - Deep and Dark Web Forensic Protocol Result compared with regular automation after Anti-forensic Bleachbit using FTK for Windows 10, AXIOM for Kali 2021.3 and UFED for ANDROID 11 and IOS 14.2.

| ARTEFACTS | Protocol | WINDOWS 10 | LINUX KALI 2021 | ANDROID 11 | APPLE IOS 14.2 |
|---|---|---|---|---|---|
| BROWSING HISTORY | Regular | 107 | 239 | 226 | 287 |
| | D2WFP | **679** | **799** | **804** | **453** |
| SECURITY & LOGINS | Regular | 07 | 09 | 06 | 02 |
| | D2WFP | **17** | **19** | **20** | **09** |
| CACHE & TEMP | Regular | 1126 | 907 | 1372 | 774 |
| | D2WFP | **4681** | **5414** | **5755** | **2413** |
| SQLITE DB FORM | Regular | 15 | 44 | 67 | 06 |
| | D2WFP | **354** | **403** | **409** | **154** |
| DOWNLOADS | Regular | 25 | 29 | 22 | 09 |
| | D2WFP | **109** | **174** | **188** | **68** |

* Results are generated automatically without accounting for empty or invalidated entries

## FORENSIC PROCESSING, EXAMINATION AND ANALYSIS

The created images need to be processed and analysed using relevant forensics software according to the evidence file types. FTK and Magnet are used for the Windows 10 images to ensure that all potential artefacts are picked up to compare the two industrial tools. To analyse the evidence files and provide comparison results, we used a Cellebrite Physical Analyser and Magnet. When processing the Windows images in FTK, we chose the most appropriate forensic profile to ensure we could find all relevant artefacts after the completion of the FTK processing. The steps for processing the evidence within AXIOM were very similar to FTK. For mobile devices, we imported the .ufdx file generated by the Cellebrite tablet to process the mobile images into the Cellebrite Physical Analyzer. In addition, we attempted processing the Caches and RAM dump obtained from the remaining devices using the AXIOM Magnet embedded Volatility module. Figure 10 illustrates the performed forensics processing, construction, reconstruction, and indexation process.

In terms of forensics processing, we analysed and examined the indexed data from both automated tools and D2WFP. We initially examined BROWSING HISTORY, which contains data about the navigation history of the user, which is used to track down if the user visited some malicious sites hosted on the dark web; we examined the data related to web searching alongside navigation history to get complete insight on the browsing activities.

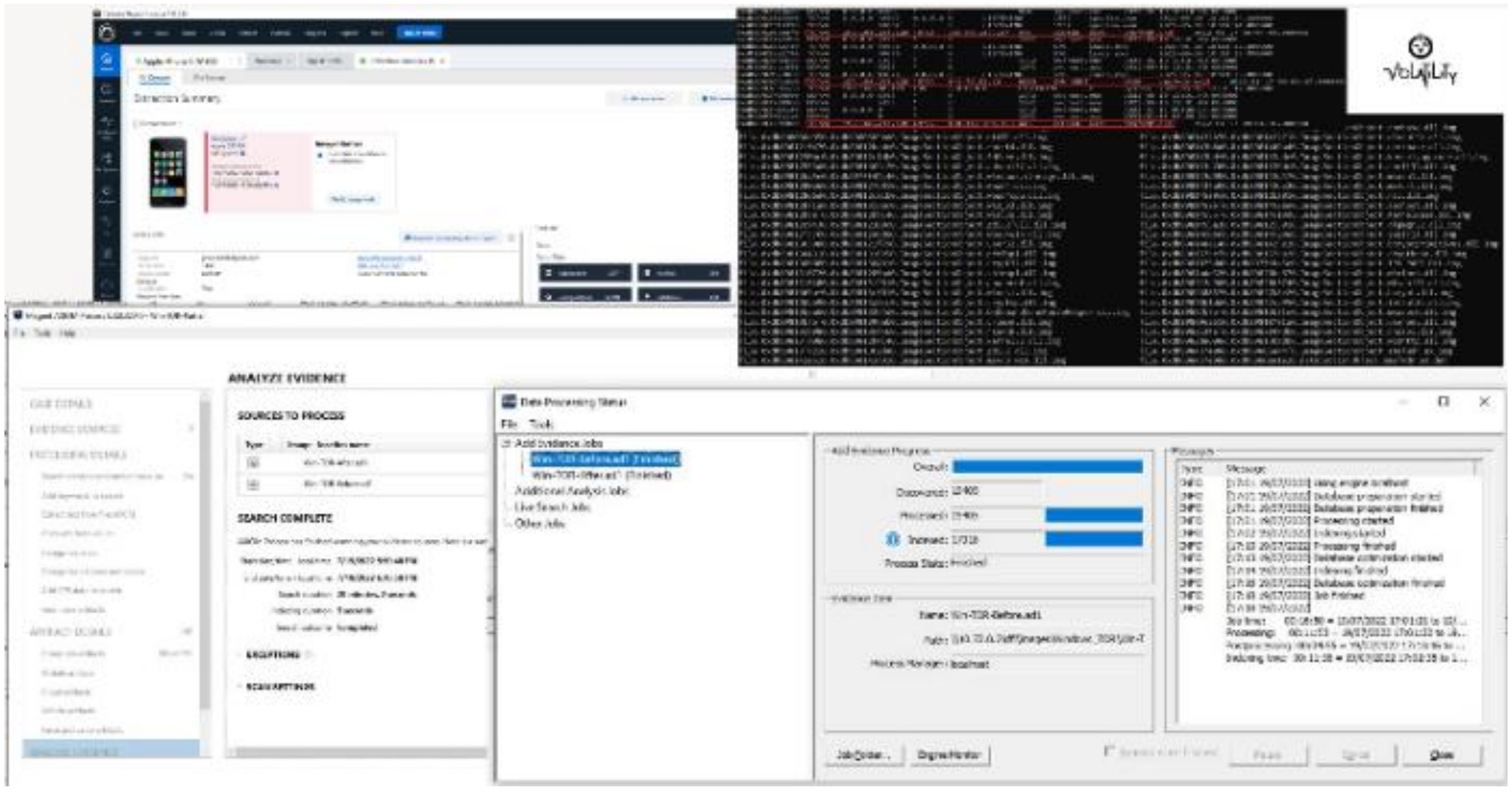

Figure 10: Forensics processing including data recovery, reconstruction and indexation using FTK, AXIOM, UFED and Volatility.

Secondly, we examined SECURITY and LOGINS data, including security logging, saved passwords, extensions, and Addons. Thirdly we examine CACHE and TEMPORARY data by looking into self-explanatory. A cache is generated when navigating websites and all sorts of temporary files created by the browser, notably cache data, images, and JavaScript, which are great data sources during a forensic investigation. Fourthly, we examined all forms and database files used or generated by the browser; this step aims to acquire more information about the website or places the user visited, including FORM DATA which is anything typed inside forms and stored by the browser. Finally, we examined the data related to data transfer and specifically DOWNLOADS.

## RESULTS AND DISCUSSION

In this section, we discuss the effectiveness of the proposed protocol D2WFP by examining and performing a quantitative evaluation which consists of comparing the results found in forensic testing of D2WFP against the results generated from automated DFIR tools and framework, namely FTK, AXIOM and UFED. The carried testing is limited to the use of anonymity and privacy preseracing browser in criminal activities and only cover artefacts recovered from criminal machines or devices. During the elaboration of the protocol, several tests were carried out, and the results obtained enabled us to make some tweaks or changes to improve D2WFP. The first set of tests covered a comparative analysis of the number of investigative leads (deep and dark web artefacts) related to web browsing using TOR. We clearly distinguish the higher number of artefacts indexed when adopting D2WFP than those indexed using the relevant tool/framework. The difference observed is significantly higher in Linux, Windows, and Android and notably in the number of BROWSING HISTORY, CACHE and TEMPORARY files. D2WFP outperforms the automated tools in all other categories, such as SECURITY, LOGINS, SQLite DB, FORM and DOWNLOADS. In the iOS context, despite having the number of indexed browsing artefacts in favour of D2WFP, the gap is less significant due to security and memory handling restrictions implemented in iOS and macOS. The analysis of the ongoing results illustrate the D2WFP effectiveness in locating, extracting, and exploiting deep and dark web activities through indexing on average 35-45% more the number of artefacts discovered by industrial tools when working in regular automation mode.

The second set of tests covered the analysis of the four forensic images obtained after running an anti-forensics software Bleach-bit to erase browsing data artefacts. Here again, the obtained results illustrate the effectiveness of D2WFP compared with regular automated DFIR tools and frameworks in terms of the number of deep and dark web artefacts related to web browsing using TOR indexed. The proportion in terms of the difference between regular automation in FTK, AXIOM and UBED compared with D2WFP is distinguished, notably the higher number for artefacts indexed in BROWSING, SECURITY and CACHE when figures show that by adopting D2WFP, the number of indexed artefacts (mostly recovered and carved) is four times higher compared with the number of artefacts indexed using the relevant tool/framework. The observed difference is significantly higher in Linux, Windows, and Android, particularly in the number of BROWSING HISTORY, CACHE and TEMPORARY files. D2WFP outperforms the automated tools in all other categories, such as SECURITY, LOGINS, SQLite DB, FORM and DOWNLOADS. In the context of iOS, the number of indexed artefacts is in favour of D2WFP, but the gap is less significant due to security.

To sum up, the purposed D2WFP is introduced here to formally structure the activities of browsing artefacts investigation and guide DFIR practitioners in tackling deep and dark web browsing artefacts investigation, the protocol applies to most of the anonymity and privacy preservation browsing tools and not only TOR which was adopted for our case study. Forensically investigating DDW browsing activities when adopting D2WFP outperform any regular forensic automated tool as the obtained results validated in comparison with FTK, AXIOM and UFED which are the industry leaders. The adoption of D2WFP resulted in finding, extracting, and reconstructing more artefacts than with those automated tools (regular automation) and the obtained results were validated in the context of the four most used OSs namely Windows, Linux, Android and iOS.

## CONCLUSIONS AND FUTURE WORKS

Cybercriminal activity on or through the dark web is on the rise. Although designed as legitimate tools for online security, anonymity, and privacy purposes, browsing browsers allow public access to dark and dark web parts that are not ordinarily searchable, indexed, or accessible through standard browsers. This research proposed a novel and comprehensive investigative protocol to guide and assist digital forensics professionals in investigating crimes committed on or via the deep and dark web. In this research, we critically analyzed the limitation in current automation, identified the research gap and developed D2WFP following scientific and experimental approaches. The proposed D2WFP protocol incorporates new and improved existing methods, mainly by establishing a sequence for performing tasks and subtasks to improve current tools' output accuracy and effectiveness, observing the order of volatility, and implementing a systemic approach covering all browsing-related hives and artefacts. A quantitative examination of the protocol capabilities was carried out following the testing using several cases, both computer and mobile devices running four different OS. The investigations and examinations were conducted using the professional version of Access Data FTK, Magnet AXIOM, and Cellebrite UFED. The results show an apparent increase in the number of artefacts recovered when adopting D2WFP compared with regular tools and framework-based automated investigations. In future work, we will consider the impact of the different levels of security in all anonymity and privacy-preserving browsers (TOR, FREENET, WATERFOX, TAILS) and will analyze the security settings' impact on the DFIR activities mainly by considering the different modes such as "standard", "safer", and "paranoid".

**Research Ethics Considerations:** Datasets were created from the forensics imaging of privately-owned Computer and Mobile devices. The deep and dark web activities simulations only covered previously reported criminal websites and were carried out with the Digital Forensics Laboratory, on factory Reset Machines and by fully qualified Certified Forensics Examiners in respect of the Computer Misuse Act 1990 and Regulation of Investigatory Powers (RIPA) Act 2000, along with Ethical principles set by Biometrics and Forensics Ethics Group, notably the Governing Principle number 5.

**Research Funding:** This work was supported in part by the UK HEIF fund from the Cyber Security Research Center under Grant RES02M3732.

**Author Contributions:** Conceptualization, MCG, PM and DD; methodology, MCG, PM and KO; software, MCG, PM, RD, DD; validation, MCG, PM, RD, DD.; formal analysis, MCG and KO; investigation, MCG, PM and DD; resources, MCG; writing—original draft preparation, MCG, PM and DD; writing—review and editing, KO and RD.; supervision, MCG and KO; project administration, KO and MCG funding acquisition, MCG.

**Data Availability Statement:** datasets generated during this research work are available to research upon request

**Conflicts of Interest:** The authors declare no conflict of interest.

## References

1. Arshad, M. R., Hussain, M., Tahir, H. & Qadir, S. (2021). Forensic Analysis of Tor Browser on Windows 10 and Android 10 Operating Systems. IEEE Access, Volume 9, pp. 141273 - 141294.
2. Balduzzi, M. & Ciancaglini, V., 2015. Cybercrime in the Deep Web. Amsterdam, 2015 Black Hat EU Conference.
3. Baronia, D., 2021. Dark Web and Tor Forensic. [Online] Available at: https://informaticss.com/dark-web-and-tor-forensic/ [Accessed 12 October 2022].
4. Brinson, R., Wimmer, H., Cheng, L. (2022). Dark Web Forensics: An investigation of tracking dark web activity with digital forensics. Interdisciplinary Research in Technology and Management (IRTM). DOI: 10.1109/IRTM54583.2022. 9791646
5. Gehl, R.W., 2018. Weaving the dark web: Legitimacy on Freenet, Tor, and I2P. MIT Press.
6. Cherty, A., Sharma, U. (2019). Memory forensic analysis for investigation of online crime- A review. IEEE 6th International Conference on Computing for Sustainable Global Development. IEEE Access.
7. European Monitoring Centre for Drugs and Drug Addiction and Europol, (2017). Drugs and the darknet: Perspectives for enforcement, research and policy, Luxembourg: Publications Office of the European Union.
8. Forensic-Pathways, 2020. Dark Web Investigations/Monitoring. [Online] Available at: https://www.forensic-pathways.com/dark-web-investigations monitoring/ [Accessed 12 October 2022].
9. Godawatte, K., Raza, M., Murtaza, M. & Saeed, A., 2019. Dark Web Along with the Dark Web Marketing and Surveillance. 2019 20th International Conference on Parallel and Distributed Computing, Applications and Technologies (PDCAT). Gold Coast, QLD, Australia, IEEE.
10. Goodison, S. E. et al. Research Report. 2019. Identifying Law Enforcement Needs for Conducting Criminal Investigations Involving Evidence on the Dark Web, California: RAND Corporation.
11. Handalage, U., Prasanga, T. (2020). Dark Web, Its Impact on the Internet and the Society: A Review. (Online) Available: DOI:10.13140/RG.2.2.11964.36484.
12. Protrka, N. (2021). Cybercrime. In Modern Police Leadership (pp. 143-155). Palgrave Macmillan, Cham.

13. R. Brinson, H. Wimmer and L. Chen, "Dark Web Forensics: An Investigation of Tracking Dark Web Activity with Digital Forensics," 2022 Interdisciplinary Research in Technology and Management (IRTM), Kolkata, India, 2022, pp. 1-8, doi: 10.1109/IRTM54583.2022.9791646.
14. M. F. B. Rafiuddin, H. Minhas and P. S. Dhubb, "A dark web story in-depth research and study conducted on the dark web based on forensic computing and security in Malaysia," *2017 IEEE International Conference on Power, Control, Signals and Instrumentation Engineering (ICPCSI)*, Chennai, India, 2017, pp. 3049-3055, doi: 10.1109/ICPCSI.2017.8392286
15. Leng, T., Yu, A. (2021). A framework of darknet forensics. International Conference on Advanced Information Science and Systems. (Online). https://dl.acm.org/doi/fullHtml/10.1145/3503047.3503082. https://doi.org/10.1145/3503047.3503082
16. Maisammaguda, D., (2019). Digital Notes on Computer Forensics, India: Malla Reddy College of Engineering and Technology. Maryville University, 2017. Top 4 Data Analysis Techniques That Create Business Value. [Online] Available at: https://online.maryville.edu/blog/data-analysis-techniques/#qualitative [Ac- cessed 4 December 2022].
17. Matic, S., Kotzias, P. and Caballero, J., 2015, October. Caronte: Detecting location leaks for deanonymizing tor hidden services. In Proceedings of the 22nd ACM SIGSAC Conference on Computer and Communications Security (pp. 1455-1466)
18. Naza, S., Huda, S., Abawajy, J., Hassan, M. M. (2020). The evolution of dark web threat and detection: a systematic approach. IEEE Access 8:171796-171819. DOI:10.1109/ACCESS.2020. 3024198.
19. Rogers, B., 2017. Tor: Beginners to Expert Guide to Accessing the DarkNet, TOR Browsing, and Remaining Anonymous Online. 1st ed: CreateSpace Independent Publishing Platform.
20. Tazi, F., Shrestha, S., Cruz, J. D. L., Das, S., (2020). SoK: An Evaluation of the Secure End User Experience on the Dark Net through Systematic Literature Review. J. Cybersecurity and Privacy. 2022, 2(2), 329-357; https://doi.org/10.3390/jcp2020018
21. Zeid, R. B., Moubarak, J., Bassil, C. (2020) Investigating the darknet (Online). International Wireless Communications and Mobile Computing (IWCMC). DOI:10.1109/IWCMC48107.2020.9148422.
22. Ozkaya, E., & Islam, R. (2019). Inside the Dark Web. CRC Press. https://doi.org/10.1201/9780367260453.
23. Popov, O., Bergman, J., & Valassi, C. (2018, November). A framework for forensically sound harvesting the dark web. Central European Cybersecurity Conference 2018 (pp. 1-7).
24. Nazah, S., Huda, S., Abawajy, J., & Hassan, M. M. (2020). Evolution of dark web threat analysis and detection: A systematic approach. IEEE Access, 8, 171796-171819.
25. Holland, B. J. (2020). Transnational cybercrime: The dark web. Encyclopedia of Criminal Activities and the Deep Web, 108-128.
26. Jardine. E. (2015). The Dark Web Dilemma: Tor, Anonymity and Online Policing. Global Commission on Internet Governance Paper Series, Volume 21, pp. 1-11. DOI:10.2139/SSRN.2667711.
27. Ghanem, M.C. Cryptographically Upgrading TOR Network to Enforce Anonymity by Enhancing Security and Improving Performances. *Preprints.org* 2023, 2023070982. https://doi.org/10.20944/preprints202307.0982.v1
28. Samtani, S., Zhu, H., & Chen, H. (2020). Proactively identifying emerging hacker threats from the dark web: A diachronic graph embedding framework. ACM Transactions on Privacy and Security (TOPS), 23(4), 1-33.
29. Dunsin, D., Ghanem, M., Ouazzane, K. (2022), 'The Use of Artificial Intelligence in Digital Forensics and Incident Response in a Constrained Environment', World Academy of Science, Engineering and Technology, Open Science Index 188, International Journal of Information and Communication Engineering, 16(8), 280 - 285.